\documentstyle[12pt]{article}
\begin{document}
\begin{center}
{\Large
\bf Transition energy and lifetime for the ground state
 hyperfine splitting of high Z lithiumlike ions}\\
\end{center}

\begin{center}
{V. M. Shabaev$^{1}$, M. B. Shabaeva$^{2}$, I. I. Tupitsyn$^{1}$,
 V. A. Yerokhin$^{3}$,
A. N. Artemyev$^{1}$,
T. K\"uhl$^{4}$,
 M. Tomaselli$^{5}$}, and
 O. M. Zherebtsov $^{1}$

\end{center}

$^{1}$ {\it Department of Physics, St.Petersburg
State University,\\ Oulianovskaya Street 1, Petrodvorets,
St.Petersburg 198904, Russia}\\
$^{2}$ {\it Department of Calculation Mathematics,\\ North-Western
Extra-Mural Polytechnical Institute,\\ Millionnaya Street 5, St.Petersburg
191065, Russia}\\
$^{3}$ {\it Institute for High Performance Computing and Data Bases,
Fontanka 118, St.Petersburg 198005, Russia}\\
$^{4}$ {\it Gesellschaft f\"ur Schwerionenforschung (GSI), Postfach 11 05 52,
D-64223 Darmstadt, Germany}\\
$^{5}$ {\it Institute f\"ur Kernphysik, Technische Hochschule Darmstadt,
Schlossgartenstrasse 9, D-64289 Darmstadt, Germany}\\

\begin{abstract}
The ground state hyperfine splitting values and the transition
probabilities between the hyperfine structure components
of high Z lithiumlike
ions are calculated in the range $Z=49-83$.
 The relativistic, nuclear, QED and
interelectronic interaction corrections
are taken into account.
It is found that the Bohr-Weisskopf
effect can be
eliminated  in a combination of the
 hyperfine splitting values of the hydrogenlike and
lithiumlike ions of an isotope. This gives
a  possibility for testing the QED effects in a
combination of the strong electric and magnetic fields
of the heavy nucleus.
Using the experimental result for the $1s$ hyperfine splitting
in  $^{209}{\rm Bi}^{82+}$ ,
 the $2s$ hyperfine splitting in   $^{209}{\rm Bi}^{80+}$
 is calculated to be
$\Delta E=0.7969(2)$ eV.
\\
$\;$
\\
PACS number(s): 31.30.Gs, 31.30.Jv
\end{abstract}

\newpage
\section{Introduction}
Recently,
laser spectroscopic measurements of the ground state hyperfine splitting
in high $Z$ hydrogenlike ions became possible at ESR [1]
and at the Super-EBIT [2]. The present status of theory of
the hyperfine splitting in high Z hydrogenlike ions is discussed in [3].
One of possible directions of further experiments is an extension
of the investigations to high $Z$ lithiumlike ions. Recently
the hyperfine structure values of lithiumlike ions were
calculated in the range $Z=7-30$ [4] in connection with
astrophysical search and for $^{209}{\rm Bi}^{80+}$ (without the QED correction)
[5] in connection with experiments at GSI [1]. In  Sec. 2
of the present paper we refine the calculation of [5] for
$^{209}{\rm Bi}^{80+}$, considering a more accurate treatment of the nuclear
effects and taking into account the QED corrections, and extend
it to lithiumlike ions in the range $Z=49-83$, which are likely
candidates for the experiments.
 In addition, a method based on
using the experimental values of the $1s$ hyperfine splitting
for determination of the
Bohr-Weisskopf effect in the lithiumlike ions is proposed.
This method is used to reduce the uncertainty of the
ground state hyperfine splitting in $^{209}{\rm Bi}^{80+}$
and $^{165}{\rm Ho}^{64+}$.
It gives a  possibility  for testing the magnetic sector
of QED.
 In Sec. 3 the transition
probabilities between the hyperfine structure components
are calculated.

\section{Hyperfine splitting values}
The energy difference between the ground state hyperfine
 splitting components of a lithiumlike ion
is conveniently written in the form [5]:

\begin{eqnarray}
\Delta E_{(1s)^{2}2s}&=&\frac{1}{6}\alpha (\alpha Z)^{3}\frac{m}{m_{p}}
\frac{\mu}{\mu_{N}}\frac{2I+1}{2I} mc^{2}\Bigl\{
[A(\alpha Z)(1-\delta)(1-\varepsilon)+x_{\rm rad}]
\nonumber\\
&&+
\frac{1}{Z}B(\alpha Z)+\frac{1}{Z^{2}}C(\alpha Z)+\cdots\Bigr\}\;.
\end{eqnarray}
Here $\alpha$ is the fine structure constant,
 $I$ is the nuclear spin, $\mu$ is the nuclear magnetic moment,
$\mu_{N}$ is the nuclear magneton, $m_{p}$ is the proton mass.
$A(\alpha Z)$ is the one-electron relativistic factor
\begin{eqnarray}
A(\alpha Z)=\frac{2[2(1+\gamma)+\sqrt{2(1+\gamma)}]}{(1+\gamma)^{2}
\gamma (4\gamma^{2}-1)}=1+\frac{17}{8}(\alpha Z)^{2}+\frac{449}{128}
(\alpha Z)^{4}+\cdots\;,
\end{eqnarray}
$\gamma=\sqrt{1-(\alpha Z)^{2}}$.  $\delta$ and $\varepsilon$ denote the
nuclear charge and magnetization distribution corrections.
 $x_{\rm rad}$ is the one-electron
radiative correction. The terms $B(\alpha Z)/Z$ and $C(\alpha Z)/Z^{2}$
correspond to interelectronic interaction contributions.

\subsection{One-electron contribution}

The one-electron contribution is enclosed in the square brackets of
the equation (1). We denote it by $a_{2s}$:
\begin{eqnarray}
a_{2s}=A^{(2s)}(\alpha Z)(1-\delta^{(2s)})(1-\varepsilon^{(2s)})
+x_{\rm rad}^{(2s)}\,.
\end{eqnarray}
To calculate the nuclear charge distribution correction $\delta$ we used
the two-parameter Fermi model
with the  parameters
taken from [6].

\subsubsection{ Bohr-Wesskopf effect}

The Bohr-Weisskopf correction $\varepsilon$  is given by
the equations
\begin{eqnarray}
\varepsilon&=&\frac{\langle IM_{I}|
\Delta Q_{S}^{z}+\Delta Q_{L}^{z}+\Delta Q_{\rm SO}^{z}|IM_{I}\rangle}
{\langle IM_{I}|Q_{\mu}^{z}|IM_{I}\rangle}\,,\\
\Delta Q_{S}^{z}&=&\sum_{i=1}^{A}g_{si}[s_{zi}K_{S}(r_{i})
+\sqrt{\frac{\pi}{2}}[Y_{2i}\otimes\sigma_{i}]^{1}_{z}
(K_{S}(r_{i})-K_{L}(r_{i}))]\,,\\
\Delta Q_{L}^{z}&=&\sum_{i=1}^{A}g_{li}l_{zi}K_{L}(r_{i})\,,\\
\Delta Q_{\rm SO}^{z}&=&\sum_{i=1}^{A}
g_{li}
\frac{2m_{p}}{3\hbar^{2}}
[s_{zi}+\sqrt{\frac{\pi}{2}}[Y_{2i}\otimes\sigma_{i}]^{1}_{z}]
\phi(r_{i})r_{i}^{2}K_{L}(r_{i})\,,\\
Q_{\mu}^{z}&=&\sum_{i=1}^{A}[g_{si}s_{zi}+g_{li}l_{zi}
+g_{li}\frac{2m_{p}}{3\hbar^{2}}
(s_{zi}+\sqrt{\frac{\pi}{2}}[Y_{2i}\otimes\sigma_{i}]^{1}_{z})
\phi(r_{i})r_{i}^{2}]\,,
\end{eqnarray}
where
$$
K_{S}(r)=\frac{\int_{0}^{r}fg\,dr'}{\int_{0}^{\infty}fg\,dr'}\;,
$$
$$
K_{L}(r)=\frac{\int_{0}^{r}\Bigl(1-\frac{r'^{3}}{r^{3}}\Bigr)
fg\,dr'}{\int_{0}^{\infty}fg\,dr'}\;,
$$
$g$ and $f$ are the radial parts of the Dirac wave function of
the electron  defined by
$$
\psi_{n\kappa m}({\bf r})=
\left(\begin{array}{c}
g_{n\kappa}(r)\Omega_{\kappa m}({\bf n})\\
if_{n\kappa}(r)\Omega_{-\kappa m}({\bf n})
\end{array}\right)\;.\\
$$
$A$ is the number of nucleons.
 The term $\Delta Q_{\rm SO}^{z}$ and the related term in (8) are caused by the
spin-orbit interaction:
$$
V_{\rm SO}(r)=\phi(r)({\bf s}\cdot{\bf l})\,.
$$
Neglecting these terms gives the equations derived in [7].
In the case of the single particle nuclear model
the equations (4)-(8) were used in [3] for calculations
of the Bohr-Weisskopf effect for the $1s$ state.
We  extended these calculations to the $2s$ state.
The uncertainty due to deviation from the single particle
nuclear model  was estimated in the same way as in [3].
This uncertainty gives a dominant contribution to the error bars of
the hyperfine splitting values. So, more accurate calculations
of the Bohr-Weisskopf effect including  a consequent procedure
for determination of the uncertainty are necessary.
Such calcualations, based on a dynamic-correlation model [8,9],
are underway and will be published elsewhere.
However, the uncertainty of the Bohr-Weisskopf effect can be
considerably reduced
 if the $1s$ hyperfine
splitting value is known from experiment with
 sufficient precision. To explain this point, let us
consider the equations (4)-(8).
As one can see from these equations, the Bohr-Weisskopf effect
depends of the electronic structure only through the functions
$K_{S}(r)$ and $K_{L}(r)$.
Simple approximate expressions for these functions
were derived in [10].
As it follows from these expressions
and is confirmed by more accurate calculations,
 with high precision ($\sim$ 0.1\% for $Z$=83)
 the functions
$K_{S}(r)$ and $K_{L}(r)$
  for
the $2s$ state are different from those for the $1s$ state
only by an overall factor denoted in [10] by $b$.
(This fact can easily be understood if we take into
account that
the binding energy of the electron ($W=E-mc^2$) is small
in comparison with the nuclear potential
 ($V(r)$) in the nuclear region. So, the binding energy
 gives only a small correction to behaviour
of the functions $g(r)$ and $f(r)$ within the nucleus.)
It follows that the Bohr-Weisskopf effect for the $2s$
state can be  found by using $\varepsilon$
for the $1s$ state and the values of the overall factors
tabulated in [10]
%(see also Appendix).
(while the
 relative precision of $b$ in [10] is of order
$\alpha Z R/(\hbar/mc)$, where $R$ is the nuclear radius,
the precision
of the ratio $b^{(2s)}/b^{(1s)}$ is  higher by orders of magnitude).
If the $1s$ hyperfine splitting is known from experiment,
the Bohr-Weisskopf effect for the $1s$ state
is derived from the equation
\begin{eqnarray}
\varepsilon^{(1s)}=\frac{\Delta E_{\rm NS}^{(1s)}
+\Delta E_{\rm QED}^{(1s)}
-\Delta E_{\rm exp}^{(1s)}}
{\Delta E_{\rm NS}^{(1s)}}\,,
\end{eqnarray}
where $\Delta E_{\rm NS}^{(1s)}$ is the theoretical
 hyperfine splitting value including the relativistic and
nuclear charge distribution effects,
$\Delta E_{\rm QED}^{(1s)}$ is the theoretical QED contribution,
and $\Delta E_{\rm exp}^{(1s)}$ is the experimental value
of the $1s$ hyperfine splitting.
The Bohr-Weisskopf effect for the $2s$ state
is calculated by
\begin{eqnarray}
\varepsilon^{(2s)}=\varepsilon^{(1s)}\frac{b^{(2s)}}{b^{(1s)}}\,.
\end{eqnarray}
High precision experimental values of the $1s$ hyperfine
splitting  were found for $^{209}{\rm Bi}^{82+}$ ($\lambda=243.87(4)$ nm)
[1] and for $^{165}{\rm Ho}^{66+}$ ($\lambda=572.79(15)$ nm) [2].
Using these experimental values and the related
 theoretical values from [3]
 (with $\mu=4.1106(2)\mu_{N}$ for $^{209}{\rm Bi}$ [11] and
$\mu=4.132(5)\mu_{N}$ for $^{165}{\rm Ho}$ [12,13,2]),
the equations (9)-(10) give
$\varepsilon^{(1s)}=0.0152(3)$,
$\varepsilon^{(2s)}=0.0164(4)$
 for $^{209}{\rm Bi}^{82+}$
 and
$\varepsilon^{(1s)}=0.0095(13)$,
$\varepsilon^{(2s)}=0.0101(14)$
for $^{165}{\rm Ho}^{66+}$. For comparison, the direct calculation,
based on the single particle nuclear model, gives
$\varepsilon^{(1s)}=0.0118(49)$,
$\varepsilon^{(2s)}=0.0127(53)$
 for $^{209}{\rm Bi}^{82+}$
 and
$\varepsilon^{(1s)}=0.0089(27)$,
$\varepsilon^{(2s)}=0.0094(28)$
for $^{165}{\rm Ho}^{66+}$.

\subsubsection{QED corrections}

The radiative correction is the sum of the vacuum polarization (VP) and
self energy (SE) contributions. The VP contribution
 is largely made up of the Uehling term.
Calculation of this term
was done in the same way as for the $1s$ state [3,14].
As to  the Wichman-Kroll
term, we calculated only the electric loop correction to the electron
wave function expecting that, like the VP screening
contribution for two-electron ions [15], the magnetic loop term is too small.

To calculate the SE contribution we used a covariant way
based on expansion of the electron propagator in terms
of external field [16,17].
The formal expression for this contribution can easily be
derived using the Green function method (see, e.g.,
[5]). The contribution of the diagram
with the hyperfine interaction outside the self-energy loop is
 divided into irreducible and reducible parts.
The reducible part is the part in which the
intermediate-state energy (between the self-energy loop and the hyperfine
interaction) coincides with the initial-state energy. The irreducible
part is the remaining one. The irreducible part is calculated in the same
way as the first order self-energy contribution.
The reducible part is grouped with the vertex diagram.
According to the Ward identity the counterterms for the vertex and the
reducible parts cancel each other and, so,
 the sum of these terms regularized in
the same covariant way is ultraviolet finite. To cancel the
 ultraviolet divergences we separate free propagators from the
bound electron lines and calculate them in the momentum
representation.  The remainder is ultraviolet finite but contains
infrared divergences, which are explicitly separated and cancelled.
The results of our calculation
for a finite nuclear charge distribution
for the $1s$ and $2s$ states,
 expressed in terms of the function
 $F(\alpha Z)$ defined by
\begin{eqnarray}
\Delta E_{\rm SE}=\frac{\alpha}{\pi}F(\alpha Z)\Delta E_{\rm NS}\,,
\end{eqnarray}
 are given in the table 1. A more detailed analysis of
the calculation is given in Ref. [18], which contains also
the results for a point nucleus.
In the case of the $1s$ state the calculation of the SE
contribution to the hyperfine splitting for a finite nuclear
charge distribution  was done first in [17,19] in a wide interval
of $Z$. In the case of $Z=83$ and a point nucleus such a calculation
was done in [20] where it was found
$x_{\rm SE}=-3.8\alpha$.
The present calculation
for $Z=83$ and a point nucleus
gives $x_{\rm SE}=-3.94\alpha$ (in the case of an extended nucleus,
$x_{\rm SE}=-3.09\alpha$). The discrepancy of the present result
with the one of Ref. [20] is caused by a spurious term which appears in
the non-covariant regularization procedure used in [20].
 A comparison of the present calculation for an extended
 nucleus with Refs. [17,19]
finds some discrepancy too.
So, for $Z=83$ our result is $F=-5.14(1)$ while in [17,19]
 it was obtained $F=-5.098$.
This discrepancy results from a small term in the vertex
contribution omitted in [17,19].
For the 1s state our results are
in good agreement with a recent calculation of Ref. [21]
where for $Z=83$ it is found $F=-5.1432$.

In addition to the nuclear charge distribution correction,
 there is also a nuclear magnetization distribution
correction to the QED effect (a combined QED - Bohr-Weisskopf effect).
This correction is expected to be negligible compared with
the uncertainty of the first order Bohr-Weisskopf effect.

Comparing the  VP and SE
contributions for the $1s$ and $2s$ states we found that,
within a few percent, they
 are related  by
\begin{eqnarray}
\frac{x_{\rm VP,SE}^{(2s)}}{A^{(2s)}}\approx\frac{x_{\rm
VP,SE}^{(1s)}}{A^{(1s)}}\,.
\end{eqnarray}
This means that, like the nuclear corrections
($\delta$ and $\varepsilon$) [10],
a dominant contribution to the QED corrections to the hyperfine
splitting arises from distances where the binding energy of the
electron is small compared with the nuclear potential.

The values of the one-electron corrections to the $2s$ hyperfine
splitting, with the Bohr-Weisskopf effect calculated within
the single particle nuclear model,
 are listed in the table 2.

\subsection{Interelectronic interaction corrections}

To find the function $B(\alpha Z)$ we have to calculate
the Feynman diagrams containing, in addition to the hyperfine
interaction line, a photon line corresponding to the
interelectronic interaction. Such a calculation for a point
nucleus with an approximate evaluation of the finite nuclear
size effect was done in [5]. In the present paper we calculate
this function  with accurate treatment of the nuclear
charge distribution effect. For that the formulas from [5]
and the finite basis set method for the Dirac equation [22-24]
are used. Like the one-electron contribution, it is convenient
to represent the function $B(\alpha Z)$ in
the form
\begin{eqnarray}
B(\alpha Z)=B_{0}(\alpha Z)(1-\delta_{B})(1-\varepsilon_{B})\,,
\end{eqnarray}
where $B_{0}(\alpha Z)$ is the point nucleus approximation of $B(\alpha Z)$,
$\delta_{B}$ is the nuclear charge distribution correction to this
function, and $\varepsilon_{B}$ is the nuclear magnetization
distribution correction. The values $B_{0}$ and $\delta_{B}$
are given in the second and third columns of the table 3. As one can see
from the tables 2 and 3, in agreement with an approximate evaluation
of the nuclear size effect for $B(\alpha Z)$ given in [5],
 the values $\delta_{B}$ are very close to the
related one-electron values $\delta$. It is natural to assume
that the nuclear magnetization correction $\varepsilon_{B}$ is also close
 to the related one-electron value $\varepsilon$ (this assumption
can be argued in the same way as the corresponding assumption for
$\delta$ in [5]).
 So, in the last
column of the table 3 we give the values $B(\alpha Z)$ corrected
by the factor $(1-\varepsilon)$.

The term $C(\alpha Z)/Z^{2}$ in equation (1) is small enough and
was estimated in the non-relativistic approximation:
\begin{eqnarray}
\frac{C(\alpha Z)}{Z^{2}}\approx \frac{C(0)}{Z^{2}}
\end{eqnarray}
The coefficient $C(0)$ was found from the CI-HF calculation of [4]
to be $C(0)=0.87\pm0.05$.

\subsection{Complete theoretical values}

In the table 4 we give the theoretical values of the energies and
wavelengths of the transition between the ground state hyperfine
splitting components of high $Z$ lithiumlike ions, based on using
the single particle nuclear model in the calculation of the
Bohr-Wesskopf effect.
The error bars given in the table are mainly defined by the uncertainty
of the Bohr-Weisskopf effect discussed in [3].
As is known (see tables 1 and 2 in [10]), the nuclear corrections
$\varepsilon$ and $\delta$ are weekly dependent functions of the the
principal quantum number $n$ for the $s$ states and, so,
cancel considerably in the the ratio of the $2s$ and $1s$ hyperfine
splitting values. It means that, if the value $\varepsilon$
is calculated in the same nuclear model for the $1s$ and $2s$
states, the ratio has a  higher precision than the individual
hyperfine splitting values. In this connection in the last column
of the table 4 we give the values $\eta$ defined by
\begin{eqnarray}
\eta=\frac{8\Delta E_{(1s)^{2}2s}}{\Delta E_{1s}}=
\frac{A^{(2s)}(1-\delta^{(2s)})(1-\varepsilon^{(2s)})
+x_{\rm rad}^{(2s)}+
\frac{B(\alpha Z)}{Z}+\frac{C(0)}{Z^{2}}}
{A^{(1s)}(1-\delta^{(1s)})(1-\varepsilon^{(1s)})
+x_{\rm rad}^{(1s)}}
\;.
\end{eqnarray}
These values can be useful for comparison experimental values
of the hyperfine splitting in the hydrogenlike and lithiumlike
ions of an isotope.
According to the equation (12) the one-electron QED corrections
are also considerably cancelled in the ratio (15) and, so, the
value $\eta$ is mainly defined by the functions $A(\alpha Z)$
and $B(\alpha Z)$.

More accurate calculations can be done
 for $^{209}{\rm Bi}^{80+}$ and
$^{165}{\rm Ho}^{64+}$  by
using the values of the Bohr-Weisskopf effect found above
from the  $1s$ hyperfine splitting experiments.
In addition, a combined interelectronic
interaction - QED correction can roughly be estimated
assuming
\begin{eqnarray}
\Delta E_{\rm int,QED}^{(2s)}\sim
 \frac{B(\alpha Z)}{Z}\frac{\Delta E_{\rm QED}^{(2s)}}
{A(\alpha Z)}\,.
\end{eqnarray}
This formula can be understood if we take into account that the
interelectronic interaction correction is mainly defined
by the direct Coulomb interaction of the $2s$ electron
with the  closed $1s$ shell [5].
Since a dominant contribution
to the QED correction arises from distances where the Coulomb
potential of the nucleus
is to be alone (see the text after the equation (12)),
the interaction of the $2s$ electron
with the spherically symmetric potential of
the closed $1s$ shell almost does not change the relative
value of the QED correction (it changes mainly the
normalization factor of the wave function
for small distances).
The precision of the estimate (16) is taken to be 50\%.
  Combining these corrections with
the other contributions from  the tables 2-4 gives
$\Delta E =$ 0.7969(2) eV ($\lambda=$ 1.5558(3)$\mu$m)
 for $^{209}{\rm Bi}^{80+}$
and $\Delta E =$ 0.3051(1) eV ($\lambda=$ 4.064(1) $\mu$m)
 for $^{165}{\rm Ho}^{64+}$.
The values of the individual contributions are given
in the table 5. It should be stressed here that the uncertainty
of the total hyperfine splitting values is not equal to the sum
of the uncertainties of the individual contributions given in the
table 5. It is caused by the fact that the total hyperfine splitting
value found in this way is stable enough in respect to possible
changes of the nuclear charge radius and the magnetic moment.
For explanation, let us represent the $2s$ hyperfine splitting
value in the form
\begin{eqnarray}
\Delta E^{(2s)}
&=&\Delta E_{\rm NS}^{(2s)}+
\Delta E_{\rm int,NS}^{(2s)}+
\nu(\Delta E_{\rm exp}^{(1s)}-
\Delta E_{\rm NS}^{(1s)})\nonumber\\
&&+
\Delta E_{\rm QED}^{(2s)}
+\Delta E_{\rm int,QED}^{(2s)}-
\nu\Delta E_{\rm QED}^{(1s)}\,,
\end{eqnarray}
where
\begin{eqnarray}
\nu=\frac{b^{(2s)}}{b^{(1s)}}\frac{\Delta E_{\rm NS}^{(2s)}+
\Delta E_{\rm int,NS}^{(2s)}}{
\Delta E_{\rm NS}^{(1s)}}\,,
\end{eqnarray}
$\Delta E_{\rm int,NS}^{(2s)}$ is the interelectronic interaction
contribution for a finite nuclear charge distribution.
(We note here that the theoretical value of
the Bohr-Weisskopf effect is eliminated
completly in (17).)
Taking, for example, a small variation of the magnetic moment
$\delta \mu$ we get
\begin{eqnarray}
\delta (\Delta E^{(2s)})&=&\frac{\delta \mu}{\mu}
\Bigl[(\Delta E_{\rm NS}^{(2s)}+\Delta E_{\rm int}^{(2s)})
\Bigl(1-\frac{b^{(2s)}}{b^{(1s)}}\Bigr)\nonumber\\
&&+\Delta E_{\rm QED}^{(2s)}+\Delta E_{\rm int,QED}^{(2s)}-
\nu\Delta E_{\rm QED}^{(1s)}\Bigr]\,.
\end{eqnarray}
Because the factor $(1-\frac{b^{(2s)}}{b^{(1s)}})$
is small (it constitutes -0.078 for $Z$=83) the ratio
$\delta(\Delta E^{(2s)})/\Delta E^{(2s)}$
 is smaller, at least,
by orders of magnitude than $\delta\mu/\mu$. Considering in the same way
a small variation of the nuclear charge
radius and taking into account that $\delta^{(1s)}$,
$\delta^{(2s)}$, and $\delta_{B}$ are considerably cancelled
in the equation (18) (e.g., in the case $Z=83$,
$(\delta^{(2s)}-\delta^{(1s)})/(\delta^{(1s)})$=0.069
and
$(\delta_{B}-\delta^{(1s)})/(\delta^{(1s)})$=0.14)
 we get a similar result.
 As to a small variation of
the $1s$ experimental hyperfine splitting value, we find
$\delta(\Delta E^{(2s)})/\Delta E^{(2s)}$=
$\delta(\Delta E_{\rm exp}^{(1s)})/\Delta E_{\rm exp}^{(1s)}$.
So, the uncertainty of the total hyperfine splitting value
is mainly defined by $\delta (\Delta E_{\rm exp}^{1s})$ and
the combined interelectronic interaction -QED term
estimated by (16).

\subsection{Testing QED effects}

One of the main objects of the investigations of the hyperfine
splitting of highly charged ions consists in testing QED effects
in  the strong electric and magnetic fields of  heavy nuclei.
As one can see from the table 5, the  QED contributions
for the $2s$ state are
larger than   the uncertainties of the hyperfine splitting values
 with the Bohr-Weisskopf effect found from the $1s$ hyperfine
splitting.
However, since the calculation of the
Bohr-Weisskopf effect includes the  QED correction of the $1s$ state,
it is natural to consider as a value derived from  QED
 the sum of the last three terms in the equation (17).
(Strictly speaking, division of the contributions
into QED and non-QED parts is not uniquely defined.
So, a part of the interelectronic
interaction contribution (the function $B(\alpha Z)$)
can  be considered  as a two-electron
QED effect [5]).
We find that the value derived from  QED  is
 0.0002(1) eV for $^{209}{\rm Bi}^{80+}$ and 0.0001 eV for $^{165}
{\rm Ho}^{64+}$.
Comparing these values with the uncertainty of the complete
theoretical values discussed in the previous subsection
(see also the table 5) we conclude
that  high precision
measurements of the ground state hyperfine splitting
 in hydrogenlike and lithiumlike ions of an isotope
   would give a  possibility for testing
QED effects in a combination of
 the strong electric and magnetic fields.

\section{Transition probabilities}
As is well known [25-27], the transition between
the hyperfine splitting components of an atomic level
is  a $M1$ transition. In  the hydrogenlike
approximation, which corresponds to the zeroth order in $1/Z$,
 in the case of one electron over
a closed shell
 the transition probability
 is given by the
formula
\begin{eqnarray}
w_{F\rightarrow F'}=\alpha \frac{\omega^{3}}{c^2}\frac{(2F'+1)(2j+1)^{3}}
{3j(j+1)}
\left\{{j\,F'\,I}\atop{F\,j\,\,1}\right\}^{2}
\Bigl[\int_{0}^{\infty}g(r)f(r)r^{3}dr\Bigr]^{2}\,,
\end{eqnarray}
where $\omega$ is the transition frequency,
$j$ is the electron
moment, $F$ and $F'$ are the total atomic moments in the
initial and final states, respectively,
$g(r)$ and $f(r)$ are the upper and lower radial components
of the hydrogenlike Dirac wave function.
For a point nucleus, using formulas from [28] one simply
finds
\begin{eqnarray}
\int_{0}^{\infty}g(r)f(r)r^{3}dr=
\frac{2\kappa \epsilon-mc^{2}}{4m c^{2}}\frac{\hbar}{mc}\,,
\end{eqnarray}
Here $\epsilon$ is the one-electron Dirac-Coulomb
energy, $\kappa=(-1)^{j+l+1/2}(j+1/2)$, and  $l$
is the  orbital electron moment.
For the $s$ states we get
\begin{eqnarray}
w_{F\rightarrow F'}=\alpha \omega^{3}\frac{\hbar^{2}}{m^{2}c^{4}}\;
\frac{4}{27}\;
\frac{I}{2I+1}\;\Bigl[\frac{2\epsilon}{mc^{2}}+1\Bigr]^{2}
\end{eqnarray}
Because the integrand in (20) is strongly decreasing function of
$r$ at $r\rightarrow 0$,
 the finite nuclear size corrections to
(21), (22) can be neglected.
To calculate the $1/Z$ interelectronic interaction
 correction to the transition probability
we used the technique developed in [29,5]. We found that this correction
 is small enough.
So, it increases $w$  by 0.23\% for $^{209}Bi^{80+}$ and
by 0.17\% for $^{165}Ho^{64+}$.
We note also that a calculation of the transition probability
for a many electron atom, including an approximate treatment
of the $1/Z$ term, can be done by the formula
\begin{eqnarray}
w_{F\rightarrow F'}=\alpha\frac{ \omega^{3}}{3c^2}(2F'+1)J(J+1)(2J+1)
\left\{{J\,F'\,I}\atop{F\,J\,\,1}\right\}^{2}\gamma^{2}(J)\,,
\end{eqnarray}
where
$$
\gamma(J)=\frac{\langle J M_{J}|\sum_{i}[{\bf r}_{i}\times
\mbox{\boldmath$\alpha$}_{i}]_{z}|J M_{J}\rangle}{M_{J}}\,,
$$
$J$ and $M_{J}$ are the total electronic moment and its projection,
respectively. Such a calculation, based on
the CI-HF method [4], confirms the exact (in $1/Z$) results.

The results of the calculation of the transition probabilities and
 the lifetimes ($\tau=1/w$), based on using the transition energies
from the table 4,
 are presented in the table 6.
According to the equations (20)-(22) the uncertainty of the
 transition probability is three times larger than the
uncertainty of the transition energy.

More accurate calculation of the transition
probability for $^{209}{\rm Bi}^{80+}$ and $^{165}{\rm Ho}^{64+}$, based
on the transition energies from the table 5,
 gives
 $w=$ 12.03(2) s$^{-1}$ ($\tau=0.0831(1)$ s) for $^{209}{\rm Bi}^{80+}$
and
 $w=$ 0.674(1) s$^{-1}$ ($\tau=1.483(2)$ s) for $^{165}{\rm Ho}^{64+}$.
The errors bars  are chosen to include the uncalculated
 terms.

\section*{Conclusion}

In the present paper we calculated the ground state hyperfine splitting
values and the transition probabilities between the
hyperfine structure components of high $Z$ lithiumlike ions.
We proposed a method which allows one to eliminate completely
 the Bohr-Weisskopf effect in
 a combination
of the hyperfine splitting values of the $1s$ and $2s$ states
and, so, gives a
 possibility for testing the QED effects.

Recently [30,31], the first experimental result for
 the ground state hyperfine splitting in lithiumlike bismuth
was reported to be $\Delta E_{exp}=0.820(26) $eV. It
agrees with the theoretical value found within
the single particle nuclear model
($\Delta E=0.800(4) $eV) as well as with the value obtained
by using the experimental result  for the $1s$ hyperfine splitting
($\Delta E=0.7969(2)$eV), although  it is close to the limit of the
error bar.

\section*{Acknowledgment}

Valuable discussions with Stefan Schneider
are gratefully acknowledged.
We thank Andrei Nefiodov for making Ref. [30]
available to us and the authors of Ref. [31] for
providing us their results before publication.
This work  was supported
in part by Grants No. 95-02-05571a (V.M.S., V.A.Y., A.N.A, and O.M.Z.)
and No. 97-02-18335 (M.B.S)
 from the Russian Foundation
for Basic Research.

\newpage
\begin{table}
\caption{ The self energy contribution to the $1s$ and $2s$
hyperfine splitting expressed in terms of the function
$F(\alpha Z)$ defined by the equation (11).
$\langle r^{2}\rangle ^{\frac{1}{2}}$ is the
root-mean-square charge radius of the nucleus [6].}
\vspace{0.5cm}
\begin{tabular}{|l|l|l|l|}\hline
$Z$&$\langle r^{2}\rangle ^{\frac{1}{2}}$&
$F^{(1s)}(\alpha Z)$&
$F^{(2s)}(\alpha Z)$\\ \hline
49&4.598&-2.629(5)&-2.58(1)\\ \hline
59&4.892&-3.293(7)&-3.28(2)\\ \hline
67&5.190&-3.856(8)&-3.89(2)\\ \hline
75&5.351&-4.470(9)&-4.57(2)\\ \hline
83&5.533&-5.141(10)&-5.32(3)\\ \hline
\end{tabular}
\end{table}

\scriptsize
\begin{table}
\caption{The one-electron contributions to
the $2s$ hyperfine splitting.
 $A(\alpha Z)$ is the
relativistic factor, $\delta$ is the nuclear charge
distribution correction, $\varepsilon$ is the Bohr-Weisskopf correction
calculated within the single particle nuclear model,
$x_{\rm VP}$ and $x_{\rm SE}$ are the vacuum polarization and
self energy corrections, respectively, and
$x_{\rm rad}$ is the total QED correction
(see equation (1)).}
\vspace{0.5cm}
\begin{tabular}{|l|c|c|c|c|c|c|c|} \hline
Ion&$A$&$\delta$&
$\varepsilon$&$x_{\rm VP}$&$x_{\rm SE}$&
$x_{\rm rad}$\\ \hline
$^{113}$In&1.3425&0.0175&0.0048&0.0034&-0.0079
&-0.0045\\ \hline
$^{121}$Sb&1.3791&0.0197&0.0053&0.0037&-0.0085&
-0.0048\\ \hline
$^{123}$Sb&1.3791&0.0197&0.0014&0.0037&-0.0085&
-0.0048\\ \hline
$^{127}$I&1.4188&0.0220&0.0054&0.0040&-0.0092&
-0.0052\\ \hline
$^{133}$Cs&1.4620&0.0245&0.0018&0.0044&-0.0099&
-0.0055\\
\hline
$^{139}$La&1.5089&0.0273&0.0026&0.0048&-0.0107&
-0.0059\\ \hline
$^{141}$Pr&1.5601&0.0304&0.0078&0.0052&-0.0115&
-0.0063\\ \hline
$^{151}$Eu&1.6770&0.0382&0.0084&0.0063&-0.0134&
-0.0071\\ \hline
$^{159}$Tb&1.7440&0.0427&0.0073&0.0069&-0.0145&
-0.0076\\ \hline
$^{165}$Ho&1.8175&0.0480&0.0094&0.0075&-0.0156&
-0.0081\\ \hline
$^{175}$Lu&1.9879&0.0607&0.0006&0.0091&-0.0183&
-0.0092\\
\hline
$^{181}$Ta&2.0871&0.0683&0.0018&0.0100&-0.0199&
-0.0098\\ \hline
$^{185}$Re&2.1973&0.0749&0.0130&0.0111&-0.0216&
-0.0104\\ \hline
$^{203}$Tl&2.6141&0.1054&0.0193&0.0152&-0.0278&
-0.0126\\ \hline
$^{205}$Tl&2.6141&0.1055&0.0193&0.0152&-0.0278&
-0.0126\\ \hline
$^{207}$Pb&2.6994&0.1120&0.0451&0.0160&-0.0291&
-0.0131\\
\hline
$^{209}$Bi&2.7904&0.1187&0.0127&0.0169&-0.0304&
-0.0135\\ \hline
\end{tabular}
\end{table}

\newpage

\begin{table}
\caption {The function $B(\alpha Z)$ defined by equation (1).
$B_{0}(\alpha Z)$ is the point nucleus value, $\delta_{B}$ is
the nuclear charge distribution correction, $B_{\rm NS}(\alpha Z)=
B_{0}(\alpha Z)(1-\delta_{B})$, and $B_{\rm NS,BW}(\alpha Z)=
B_{0}(\alpha Z)(1-\delta_{B})(1-\varepsilon)$. }
\footnotesize
\vspace{0.5cm}
\begin{tabular}{|c|c|c|c|c|l|} \hline
Ion&$B_{0}(\alpha Z)$&$\delta_{B}$&$
B_{NS}(\alpha Z)$&$B_{NS,BW}(\alpha Z)$ \\ \hline
$^{113}$In$^{46+}$&-3.677&0.018&-3.611&-3.594\\ \hline
$^{121}$Sb$^{48+}$&-3.788&0.020&-3.712&-3.692\\ \hline
$^{123}$Sb$^{48+}$&-3.788&0.020&-3.712&-3.707\\ \hline
$^{127}$I$^{50+}$&-3.909&0.023&-3.821&-3.800\\ \hline
$^{133}$Cs$^{52+}$&-4.042&0.025&-3.939&-3.932\\ \hline
$^{139}$La$^{54+}$&-4.186&0.028&-4.068&-4.057\\ \hline
$^{141}$Pr$^{56+}$&-4.344&0.032&-4.207&-4.174\\ \hline
$^{151}$Eu$^{60+}$&-4.707&0.040&-4.520&-4.482\\ \hline
$^{159}$Tb$^{62+}$&-4.916&0.045&-4.697&-4.663\\ \hline
$^{165}$Ho$^{64+}$&-5.147&0.050&-4.889&-4.843\\ \hline
$^{175}$Lu$^{68+}$&-5.687&0.064&-5.324&-5.321\\ \hline
$^{181}$Ta$^{70+}$&-6.003&0.072&-5.572&-5.561\\ \hline
$^{185}$Re$^{72+}$&-6.357&0.079&-5.855&-5.779\\ \hline
$^{203}$Tl$^{78+}$&-7.711&0.112&-6.849&-6.717\\ \hline
$^{205}$Tl$^{78+}$&-7.711&0.112&-6.849&-6.717\\ \hline
$^{207}$Pb$^{79+}$&-7.992&0.119&-7.042&-6.724\\ \hline
$^{209}$Bi$^{80+}$&-8.292&0.126&-7.247&-7.154\\ \hline
\end{tabular}
\end{table}
\newpage
\scriptsize
\begin{table}
\caption{The energies ($\Delta E$) and the wavelengths
($\lambda $) of the transition between
the hyperfine
structure components of the ground state of  lithiumlike ions,
with the Bohr-Weisskopf effect calculated within the single particle
nuclear model.
$a_{2s}$ is the
total one-electron contribution defined by (3).
$B(\alpha Z)/Z$ and $C(0)/Z^2$ are
the interelectronic interaction
 contributions
defined by the equation (1). $\eta=8\Delta E_{(1s)^{2} 2s}/
\Delta E_{1s}$. The nuclear magnetic moments are taken from [11-13].}
\vspace{0.5cm}
\begin{tabular}{|c|l|c|c|c|l|l|c|} \hline
Ion&$\frac{\mu}{\mu_{N}}$&$a^{(2s)}$&
$\frac{B(\alpha Z)}{Z}$&$\frac{C(0)}{Z^2}$&$\Delta E $ (eV)&$
\lambda
(\mu m)$&
$\eta$
\\ \hline
$^{113}$In$^{46+}$&5.5289(2)&1.3081&-0.0733&0.0004&0.11742(18)&10.56(2)
&1.0268(2)\\ \hline
$^{121}$Sb$^{48+}$&3.3634(3)&1.3400&-0.0724&0.0003&0.08929(15)&13.89(2)&1.0366(2)\\ \hline
$^{123}$Sb$^{48+}$&2.5498(2)&1.3452&-0.0727&0.0003&0.06472(9)&19.16(3)&1.0366(2)\\ \hline
$^{127}$I$^{50+}$&2.81327(8)&1.3749&-0.0717&0.0003&0.08617(15)&14.39(2)&1.0465(2)\\ \hline
$^{133}$Cs$^{52+}$&2.58202&1.4180&-0.0715&0.0003&0.08697(14)&14.26(2)&1.0570(2)\\
\hline
$^{139}$La$^{54+}$&2.78305&1.4580&-0.0712&0.0003&0.10746(20)&11.54(2)&1.0677(3)\\ \hline
$^{141}$Pr$^{56+}$&4.2754(5)&1.4945&-0.0707&0.0002&0.1974(5)&6.282(16)&1.0786(3)\\ \hline
$^{151}$Eu$^{60+}$&3.4717(6)&1.5922&-0.0711&0.0002&0.2084(6)&5.948(16)&1.1021(3)\\ \hline
$^{159}$Tb$^{62+}$&2.014(4)&1.6497&-0.0717&0.0002&0.1531(5)&8.10(3)&1.1145(3)\\ \hline
$^{165}$Ho$^{64+}$&4.132(5)&1.7059&-0.0723&0.0002&0.3052(10)&4.062(13)&1.1274(3)\\ \hline
$^{175}$Lu$^{68+}$&2.2327(11)&1.8569&-0.0749&0.0002&0.2141(7)&5.79(2)&1.1556(3)\\
\hline
$^{181}$Ta$^{70+}$&2.3705(7)&1.9312&-0.0762&0.0002&0.2572(9)&4.82(2)&1.1703(3)\\ \hline
$^{185}$Re$^{72+}$&3.1871(3)&1.9958&-0.0770&0.0002&0.407(2)&3.045(13)&1.1850(4)\\ \hline
$^{203}$Tl$^{78+}$&1.62226&2.2808&-0.0829&0.0001&0.498(3)&2.487(15)&1.2350(6)\\ \hline
$^{205}$Tl$^{78+}$&1.63821&2.2806&-0.0829&0.0001&0.503(3)&2.463(15)&1.2350(6)\\ \hline
$^{207}$Pb$^{79+}$&0.592583(9)&2.2759&-0.0820&0.0001&0.1886(9)&6.57(3)&1.2417(5)\\
\hline
$^{209}$Bi$^{80+}$&4.1106(2)&2.4144&-0.0862&0.0001&0.800(4)&1.550(9)&1.2542(5)\\ \hline
\end{tabular}
\end{table}
\small
\begin{table}
\caption{ The individual contributions  to the ground state hyperfine
splitting
in $^{209}{\rm Bi}^{80+}$
for
$\Delta E_{\rm exp}^{(1s)}=5.0840(8)$ eV [1],
 $\mu=4.1106(2)\mu_{N}$ [11]
 and
in $^{165}{\rm Ho}^{64+}$
for
$\Delta E_{\rm exp}^{(1s)}=2.1645(6)$ eV [2],
$\mu=4.132(5)\mu_{N}$ [12,13,2].
The Bohr-Weisskopf effect  is found by using the
experimental values of the $1s$ hyperfine splitting (see the text).}
\vspace{0.5cm}
\begin{tabular}{|l|l|l|} \hline
Contribution & $^{209}{\rm Bi}^{80+}$&$^{165}{\rm Ho}^{64+}$\\ \hline
Nonrelativistic one-electron value  &     0.34349(2)eV  & 0.1868(2)eV\\ \hline
Relativistic one-electron value & 0.95850(5) eV &0.3395(4) eV\\ \hline
Nuclear size  &       -0.1138(3) eV &-0.0163(1) eV\\ \hline
Bohr-Weisskopf   &     -0.0139(3) eV& -0.0032(5) eV\\ \hline
One-electron QED   &      -0.0046 eV&-0.0015 eV\\ \hline
Interelectronic interaction  &      -0.02945(4) eV&-0.01346(4) eV\\ \hline
Interelectronic interaction-QED & 0.00016(8) eV&0.00006(3) eV\\ \hline
Total  value &    0.7969(2) eV  & 0.3051(1) eV\\ \hline
\end{tabular}
\end{table}

\newpage
\scriptsize
\begin{table}
\caption{The transition probabilities ($w$) and the lifetimes
($\tau=1/w$) for the ground state hyperfine splitting of high $Z$
lithiumlike ions calculated with the transition energies from
the table 4.}
\vspace{0.5cm}
\begin{tabular}{|c|l|l|} \hline
Ion&$w$ (s$^{-1}$)&
$\tau$ (s)
\\ \hline
$^{113}{\rm In}^{46+}$&0.0404(2)&24.77(11)\\ \hline
$^{121}{\rm Sb}^{48+}$&0.01641(8)&61.0(3)\\ \hline
$^{123}{\rm Sb}^{48+}$&0.00656(3)&152.4(6)\\ \hline
$^{127}{\rm I}^{50+}$&0.01472(8)&67.9(4)\\ \hline
$^{133}{\rm Cs}^{52+}$&0.01585(8)&63.1(3)\\
\hline
$^{139}{\rm La}^{54+}$&0.02985(16)&33.5(2)\\ \hline
$^{141}{\rm Pr}^{56+}$&0.1757(13)&5.69(4)\\ \hline
$^{151}{\rm Eu}^{60+}$&0.206(2)&4.86(4)\\ \hline
$^{159}{\rm Tb}^{62+}$&0.0732(7)&13.66(13)\\ \hline
$^{165}{\rm Ho}^{64+}$&0.675(7)&1.481(14)\\ \hline
$^{175}{\rm Lu}^{68+}$&0.232(2)&4.32(4)\\
\hline
$^{181}{\rm Ta}^{70+}$&0.400(4)&2.50(3)\\ \hline
$^{185}{\rm Re}^{72+}$&1.51(2)&0.663(8)\\ \hline
$^{203}{\rm Tl}^{78+}$&1.64(3)&0.609(11)\\ \hline
$^{205}{\rm Tl}^{78+}$&1.69(3)&0.591(11)\\ \hline
$^{207}{\rm Pb}^{79+}$&0.0888(13)&11.27(17)\\
\hline
$^{209}{\rm Bi}^{80+}$&12.2(2)&0.0822(14)\\ \hline
\end{tabular}
\end{table}

\end{document}